\documentclass[aps,prl,superscriptaddress,, nofootinbib,twocolumn]{revtex4-1}

\usepackage{epsfig}
\usepackage{color}
\usepackage[latin1]{inputenc}
\usepackage{float,amsmath}
\usepackage{graphicx}

\newcommand{\be}{\begin{eqnarray}}
\newcommand{\ee}{\end{eqnarray}}

\newcommand{\benum}{\begin{enumerate}}
\newcommand{\eenum}{\end{enumerate}}

\begin{document}

\title{Anisotropic flow in $\sqrt{s}=2.76\,{\rm TeV}$ Pb+Pb collisions at the LHC}

\author{Bj\"orn Schenke}
\affiliation{Physics Department, Bldg. 510A, Brookhaven National Laboratory, Upton, NY 11973, USA}

\author{Sangyong Jeon}
\affiliation{Department of Physics, McGill University, 3600 University Street, Montreal, Quebec, H3A\,2T8, Canada}

\author{Charles Gale}
\affiliation{Department of Physics, McGill University, 3600 University Street, Montreal, Quebec, H3A\,2T8, Canada}

\begin{abstract}
The results on elliptic flow in $\sqrt{s}=2.76\,{\rm TeV}$ Pb+Pb collisions at the Large Hadron Collider (LHC) reported by the ALICE collaboration
are remarkably similar to those for $\sqrt{s}=200\,{\rm GeV}$ gold-gold collisions at the Relativistic Heavy Ion Collider (RHIC). This result is surprising, given the expected longer lifetime of the system at the higher collision energies. We show that it is nevertheless consistent with 3+1 dimensional viscous event-by-event hydrodynamic calculations, and demonstrate that elliptic flow at both RHIC and LHC is built up mostly within the first $\sim 5\,{\rm fm}/c$ of the evolution. We conclude that an ``almost perfect liquid'' is produced in heavy-ion collisions at the LHC. Furthermore, we present predictions for triangular flow as a function of transverse momentum for different centralities.
\end{abstract}

\maketitle

\hyphenation{ALICE}
The LHC era has barely begun, yet it is already producing significant physics
results. In particular, the ALICE collaboration has demonstrated that the QGP system
produced at $\sqrt{s} = 2.76\,{\rm TeV}$ at the LHC
is in many ways similar to the system produced by RHIC at
the much lower $\sqrt{s} = 200\,{\rm GeV}$. Specifically, the elliptic flow, measured
by the ALICE collaboration \cite{Aamodt:2010pa} is surprisingly close to that
measured by the STAR collaboration. 
Naively, one would have expected that the higher pressure and the longer lifetime
of the QGP would make the effect of spatial anisotropy much greater at the LHC.
To determine whether the LHC data 
is showing truly unexpected feature requires thorough analysis of hydrodynamics
with energy and entropy density appropriate for the LHC.
In this work, we use a 3+1D viscous hydrodynamic simulation model \cite{Schenke:2010nt,Schenke:2010rr}
to show that this is actually a natural consequence of the self-quenching of
elliptic flow when the spatial eccentricity comes down below 0.1.

As has been extensively reviewed \cite{Huovinen:2003fa,Kolb:2003dz}, the elliptic flow, defined
by the coefficient of $\cos 2\phi$ in the momentum distribution as follows 
\be
{dN\over dy d^2 p_T} 
= {1\over 2\pi}{dN\over dy p_T dp_T}\left(1 + 2v_2\cos 2\phi + \cdots\right)
\ee
has been one of the most important evidence of the quark-gluon plasma (QGP) 
\cite{Adams:2004bi,Adare:2010ux,Back:2004mh,Sanders:2007th}
.
A significant amount of theoretical work has been carried out by various groups 
for RHIC \cite{Heinz:2009xj} and some for LHC \cite{Hirano:2010je,Luzum:2010ag,Bozek:2011wa}. 
Recently, we have emphasized the role of event-by-event fluctuations of the initial condition and the
finite viscosity in a full 3+1D hydrodynamic calculation \cite{Schenke:2010rr} in
understanding the details of elliptic flow measured at RHIC. 
In this study we demonstrate the importance of both at LHC energies.

In this work, we use a variant of the Israel-Stewart formalism \cite{Israel:1976tn,Stewart:1977,Israel:1979wp,Grmela:1997zz,Muronga:2001zk}
derived in \cite{Baier:2007ix}, where the stress-energy tensor is decomposed as
\be\label{tmunu}
{\cal T}^{\mu\nu} = T^{\mu\nu}_{\rm id} + W^{\mu\nu}\,,
\ee
where
\be
T_{\rm id}^{\mu\nu} = (\epsilon + {\cal P})u^\mu u^\nu - {\cal P}g^{\mu\nu}
\ee
is the ideal fluid part with flow velocity $u^\mu$. The local energy density
and local pressure are $\epsilon$  and ${\cal P}$, respectively, and $g^{\mu\nu} = \hbox{diag}(1, -1, -1, -1)$ is the metric tensor.
The flow velocity is defined as
the time-like eigenvector of $T^{\mu\nu}_{\rm id}$
\be
T^{\mu\nu}_{\rm id} u_{\nu} = \epsilon u^\mu
\ee
with the normalization $u^\nu u_\nu = 1$.
The pressure is determined by the equation of state
\begin{eqnarray}
\mathcal{P} = \mathcal{P}(\varepsilon, \rho_B)\,.
\end{eqnarray}

The evolution equations are
$\partial_\mu {\cal T}^{\mu\nu} = 0$
and
\begin{equation}
\Delta^{\mu}_{\alpha}\Delta^{\nu}_{\beta}
{u^\sigma\partial_\sigma} W^{\alpha\beta}
=
-{1\over \tau_\pi}
\left( W^{\mu\nu} - S^{\mu\nu} \right) - {4\over 3}W^{\mu\nu}(\partial_\alpha u^\alpha)\,,
\label{eq:Weq}
\end{equation}
where
\be
S^{\mu\nu} = \eta 
\left(
\nabla^\mu u^\nu + \nabla^\nu u^\mu - 
{2\over 3}\Delta^{\mu\nu}\nabla_\alpha u^\alpha
\right)\,,
\ee
with $\Delta^{\mu\nu} = g^{\mu\nu} - u^\mu u^\nu$ the local 3-metric, $\nabla^\mu = \Delta^{\mu\nu}\partial_\nu$ the local space derivative, and $\eta$ the shear viscosity.

In the $\tau, \eta_s$ coordinate system used in this work, these equations can be
re-written as hyperbolic equations with sources
\be
\partial_a T_{\rm id}^{ab} = -\partial_a W^{ab} + F^b
\label{eq:Tid_eq}
\ee
and
\be
\partial_a (u^a W^{cd}) = -(1/\tau_\pi)(W^{cd} - S^{cd}) + G^{cd}
\label{eq:uW_eq}
\ee
where $F^b$ and $G^{cd}$ contain terms introduced by the coordinate change
from $t,z$ to $\tau,\eta_s$ as well as those introduced by the projections in
Eq.(\ref{eq:Weq}).

To solve this system of equations, the 3+1D relativistic hydrodynamic simulation \textsc{music} \cite{Schenke:2010nt,Schenke:2010rr} is used.
This approach utilizes the Kurganov-Tadmor (KT) scheme \cite{Kurganov:2000,Naidoo:2004}, together
with Heun's method to solve resulting ordinary differential equations.

The initialization of the energy density is done using the Glauber model (see \cite{Miller:2007ri} and references therein):
Before the collision the density distribution of the two nuclei is described by
a Woods-Saxon parametrization
\begin{equation}
  \rho_A(r) = \frac{\rho_0}{1+\exp[(r-R)/d]}\,,
\end{equation}
with $R=6.62\,{\rm fm}$ and $d=0.546\,{\rm fm}$ for Pb nuclei.
The normalization factor $\rho_0$ is set to fulfil $\int d^3r \rho_A(r)=A$.
The relevant quantity for the following considerations is the \emph{nuclear thickness function}
\begin{equation}
  T_A(x,y) = \int_{-\infty}^{\infty}dz\,\rho_A(x,y,z)\,,
\end{equation}
where $r=\sqrt{x^2+y^2+z^2}$.
The opacity of the nucleus is obtained by multiplying the thickness function with the total inelastic cross-section $\sigma$ of
a nucleus-nucleus collision.

The initial energy density distribution in the transverse plane is scaled 
with the number of wounded nucleons $n_{\rm WN}$. The initial energy density at the center is $175\,{\rm GeV}/{\rm fm}^3$ in 
the ideal case, and $158\,{\rm GeV}/{\rm fm}^3$ for $\eta/s=0.08$: the ratio of shear viscosity to entropy density.
For the event-by-event simulation, the same procedure as described in \cite{Schenke:2010rr} is followed.
For every wounded nucleon, a contribution to the energy density with Gaussian shape (in $x$ and $y$) and width $\sigma_0=0.3\,{\rm fm}$ is added.
The amplitude of the Gaussian was adjusted to yield the same average multiplicity distribution as in the case with average initial conditions. 
We tested the effect of partial scaling with binary collisions and found that it has very little effect on the flow observables
in the event-by-event calculation. The reason for this is the appearance of ``hot spots'' that work against the increase in initial eccentricity, because their expansion is not necessarily aligned with the event-plane.


The prescription described in Refs.  
\cite{Ishii:1992xi,Morita:1999vj,Hirano:2001yi,Hirano:2001eu,Hirano:2002ds,Morita:2002av,Nonaka:2006yn} is used to initialize the longitudinal profile, both for average initial conditions and for event-by-event simulations.
It consists of two parts, namely a flat region around $\eta_s=0$ and half a Gaussian in the forward
and backward direction:
\begin{equation}
  H(\eta_s)=\exp\left[-\frac{(|\eta_s|-\eta_{\rm flat}/2)^2}{2 \sigma_{\eta}^2}\theta(|\eta_s|-\eta_{\rm flat}/2)\right]\,.
\end{equation}
The full energy density distribution is then given by
\begin{equation}
  \varepsilon(x,y,\eta_s,b) = \varepsilon_0\, \, H(\eta_s)\,n_{\rm WN}(x,y,b)/n_{\rm WN}(0,0,0)\,.
\end{equation} 
For the LHC scenario, the parameters $\eta_{\rm flat}$ and $\sigma_\eta$ are respectively set to 10 and 0.5, in order to reproduce predictions in
\cite{Jeon:2003nx}. All parameters for RHIC are the same as in parameter set Au-Au-1 in \cite{Schenke:2010nt}.

The equation of state used in this work is that of the parametrization ``s95p-v1'' from Ref. \cite{Huovinen:2009yb}, obtained from interpolating between
lattice data and a hadron resonance gas.

A Cooper-Frye freeze-out is performed, using
\begin{equation}\label{cf}
E\frac{dN}{d^3p}=\frac{dN}{dy p_T dp_T d\phi_p} = g_i \int_\Sigma f(u^\mu p_\mu) p^\mu d^3\Sigma_\mu\,,
\end{equation}
where $g_i$ is the degeneracy of particle species $i$, and $\Sigma$ the freeze-out hyper-surface.
The distribution function is given by
\begin{equation}
  f(u^\mu p_\mu) = f_0(u^\mu p_\mu) = \frac{1}{(2\pi)^3}\frac{1}{\exp((u^\mu p_\mu -\mu_i)/T_{\rm FO})\pm 1}\,,
\end{equation}
where $\mu_i$ is the chemical potential for particle species $i$ and $T_{\rm FO}$ is the freeze-out temperature.
In the finite viscosity case we include viscous corrections to the distribution function, $f=f_0+\delta f$, with
\begin{equation}
 \delta f = f_0 (1\pm f_0) p^\alpha p^\beta W_{\alpha\beta} \frac{1}{2 (\epsilon+\mathcal{P}) T^2}\,,
\end{equation}
where $W$ is the viscous correction introduced in Eq. (\ref{tmunu}).
Note however that the choice $\delta f \sim p^2$ is not unique \cite{Dusling:2009df}.


The algorithm used to determine the freeze-out surface $\Sigma$ has been presented in \cite{Schenke:2010nt}. 
In the case with average initial conditions, we include all resonances up to $2\,{\rm GeV}$, performing resonance decays using routines 
from \cite{Sollfrank:1990qz,Sollfrank:1991xm,Kolb:2000sd,Kolb:2002ve} that we generalized to three dimensions.
For the event-by-event simulations we only include resonances up to the $\phi$-meson. We have verified that the effect of
including higher resonances on $v_2$ is negligible in the case of average initial conditions, such that neglecting them in the event-by-event case
is not expected to change the results for flow observables. The thermalization time used for the conditions at the LHC is $\tau_0=0.3\,{\rm fm}/c$ (compared to $\tau_0=0.4\,{\rm fm}/c$ for RHIC).

\begin{figure}[tb]
   \begin{center}
     \includegraphics[width=8.5cm]{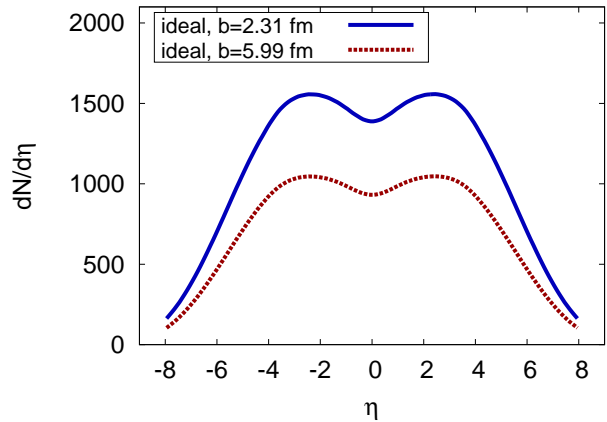}
     \caption{(Color online) Charged hadron multiplicity as a function of the pseudo-rapidity $\eta$ for most central (0-5\%) and mid-central (10-20\%)
     collisions, using $T_{FO}=136\,{\rm MeV}$.}
     \label{fig:lhceta}
   \end{center}
\end{figure}

For the sake of reference, the pseudo-rapidity dependence of the charged particle multiplicity distribution is shown in Fig.\,\ref{fig:lhceta}.
The spectrum was obtained by fitting to predictions in \cite{Jeon:2003nx} and normalizing the multiplicity at $\eta=0$ to experimental results from
\cite{Aamodt:2010pb}.

In the event-by-event simulation the same method as in \cite{Schenke:2010rr} is applied, where the flow coefficients
\begin{equation}
  v_n=\langle \cos(n(\phi-\psi_n)) \rangle\,,
\end{equation}
are measured with respect to the event plane, defined by the angle
\begin{equation}
  \psi_n=\frac{1}{n}\arctan\frac{\langle p_T \sin(n\phi)\rangle}{\langle p_T \cos(n\phi)\rangle}\,.
\end{equation}
The weight $p_T$ is chosen for best accuracy \cite{Poskanzer:1998yz}.

\begin{figure}[tb]
   \begin{center}
     \includegraphics[width=8.5cm]{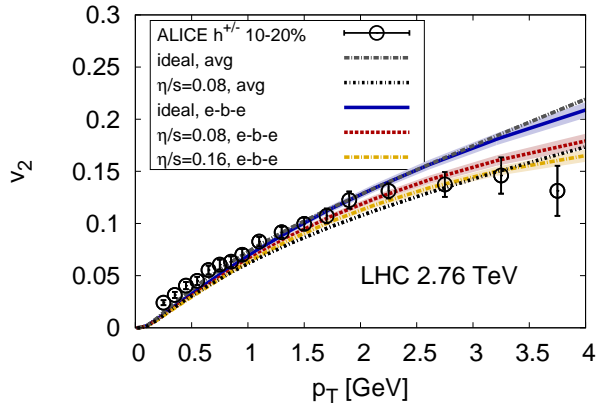}
     \caption{(Color online) Charged hadron $v_2$ in 10-20\% central collisions for average initial conditions (avg) and event-by-event simulations (e-b-e). The bands indicate the statistical error. Experimental data from ALICE \cite{Aamodt:2010pa}.}
     \label{fig:v2-10-20-new}
   \end{center}
\end{figure}

\begin{figure}[tb]
   \begin{center}
     \includegraphics[width=8.5cm]{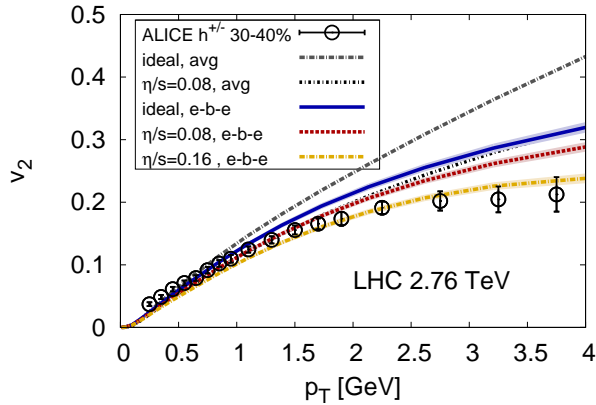}
     \caption{(Color online) Charged hadron $v_2$ in 30-40\% central collisions for average initial conditions (avg) and event-by-event simulations (e-b-e). The bands indicate the statistical error. Experimental data from ALICE \cite{Aamodt:2010pa}.}
     \label{fig:v2-30-40-new}
   \end{center}
\end{figure}

In Figs.\,\ref{fig:v2-10-20-new} and \ref{fig:v2-30-40-new} we present the elliptic flow $v_2$ as a function of transverse momentum 
$p_T$ obtained in the event-by-event simulation, compared to the average initial condition case and data from ALICE \cite{Aamodt:2010pa}. 

The used freeze-out temperature $T_{FO}=135\,{\rm MeV}$ is the same as that used to produce Fig.\,\ref{fig:lhceta}. 
Event-by-event fluctuations have opposite effects in the two different centrality bins, increasing $v_2$ in more central collisions, 
but decreasing it in 30-40\% central collisions. This has been observed and explained previously at RHIC energies in \cite{Schenke:2010rr}.
Overall, event-by-event fluctuations and finite viscosity improve the agreement with the experimental data for the 30-40\% central bin.
For $p_T<0.7\,GeV$ the experimental data is underestimated. A similar mismatch to experimental data was found in 
\cite{Hirano:2010je,Bozek:2011wa}, where \cite{Bozek:2011wa} explains the low $p_T$ $v_2$ by including non-thermalized particles from jet fragmentation.
For 10-20\% central collision the underestimation of $v_2$ at low $p_T$ is even larger. Note that we use the event-plane method to determine $v_2$, while the ALICE data is obtained using the four particle cumulant method. This method eliminates most of the non-flow contributions but also minimizes effects of fluctuations \cite{Borghini:2001vi}.  Using the event-plane method or the two-particle cumulant leads to about a 10\% larger $v_2$ \cite{Lacey:2010ej}.
Potentially, initial conditions using the Monte-Carlo KLN Model \cite{Drescher:2006ca,Drescher:2007ax}, which increase the effective initial eccentricity,
can improve the agreement with experimental data by leading to larger $v_2$.

\begin{figure}[tb]
   \begin{center}
\vspace{-0.4cm}
     \includegraphics[width=8.5cm]{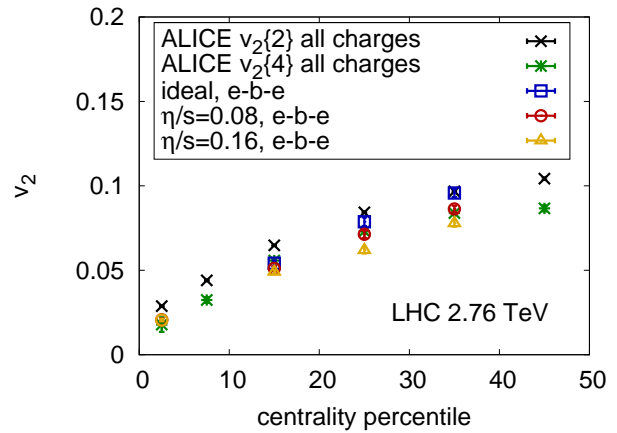}
\vspace{-0.3cm}
     \caption{(Color online) Charged hadron $v_2$ from event-by-event simulations as a function of centrality.}
     \label{fig:v2cent}
   \end{center}
\vspace{-0.7cm}
\end{figure}

Fig.\,\ref{fig:v2cent} shows $v_2$ as a function of centrality compared to two particle cumulant $v_2\{2\}$ and four particle cumulant $v_2\{4\}$ from ALICE \cite{Aamodt:2010pa}. The results reflect the slighlty too low $v_2$ at low $p_T$ seen in Figs.\,\ref{fig:v2-10-20-new} and \ref{fig:v2-30-40-new}.

\begin{figure}[tb]
   \begin{center}
     \includegraphics[width=8.5cm]{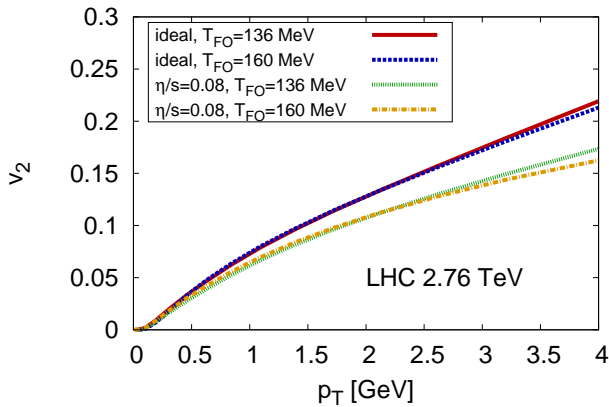}
     \caption{(Color online) Charged hadron $v_2$ as function of transverse momentum $p_T$ for the final freeze-out temperature $T_{FO}=136\,{\rm MeV}$ and
     a freeze-out at the temperature below the cross-over region $T_{FO}=160\,{\rm MeV}$. The difference is mini\-mal, indicating that all the elliptic
     flow builds up during the QGP-phase.}
     \label{fig:v2}
   \end{center}
\end{figure}

The flow coefficient $v_2$ for average initial conditions and a freeze-out temperature of $T_{FO}=160\,{\rm MeV}$ - just below the range of the cross-over transition - is shown in Fig.\,\ref{fig:v2}. It is very similar to the result obtained with $T_{FO}=137\,{\rm MeV}$, which 
demonstrates that the elliptic flow is completely built up already at the earlier time,
explaining the small difference between the $v_2(p_T)$ at RHIC and LHC.

\begin{figure}[tb]
   \begin{center}
     \includegraphics[width=8.5cm]{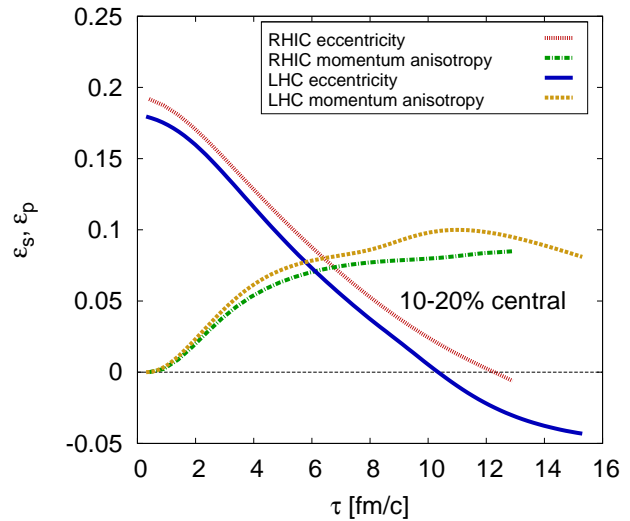}
     \caption{(Color online) Time evolution of the eccentricity and momentum anisotropy in 10-20\% central collisions at RHIC and the LHC.
     In both cases the maximal momentum anisotropy is built up almost entirely during the first $5-6\,{\rm fm}/c$. 
     At LHC the spatial anisotropy drops even slightly faster. This demonstrates why the elliptic flow as a function of $p_T$ is very similar in both experiments.}
     \label{fig:ecc}
   \end{center}
\end{figure}

\begin{figure}[tb]
   \begin{center}
     \includegraphics[width=8.5cm]{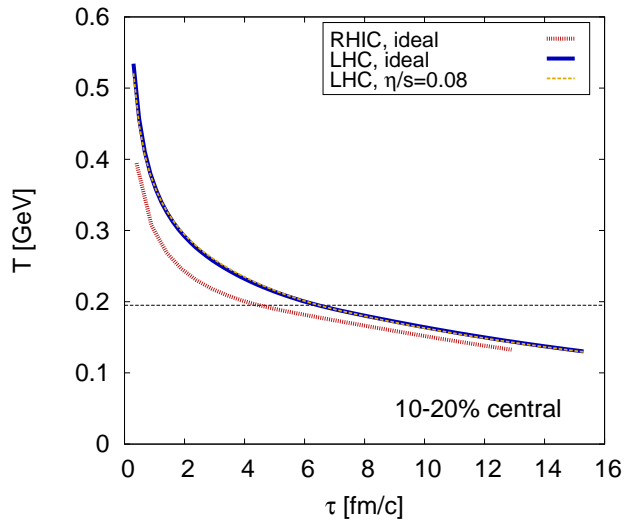}
     \caption{(Color online) Time evolution of the temperature in the center at RHIC and LHC. 
       The horizontal line indicates a temperature of $195\,{\rm GeV}$,
       which is in the center of the cross-over region defined in \cite{Huovinen:2009yb}.}
     \label{fig:T}
   \end{center}
\end{figure}

To make this point more clear, the time-dependent eccentricity
\begin{equation}
  \epsilon_s = \frac{\langle y^2-x^2\rangle}{\langle y^2+x^2 \rangle}\,,
\end{equation}
and momentum anisotropy of the system at midrapidity
\begin{equation}
  \epsilon_p = \frac{\langle T^{xx}-T^{yy}\rangle}{\langle T^{xx}+T^{yy} \rangle}\,
\end{equation}
are shown in Fig.\,\ref{fig:ecc}. 
This illustrates how the momentum anisotropy is almost entirely built up after $5\,{\rm fm}/c$ at both RHIC and LHC and the (slightly) 
longer lifetime at LHC does not play a role for $v_2$. We also find that while the system becomes just isotropic at freeze-out at RHIC,
the eccentricity changes sign approximately $5\,{\rm fm}/c$ before freeze-out at LHC.
Turning to the temperature evolution in the center of the system in Fig.\,\ref{fig:T},
one concludes that at the LHC, almost all elliptic flow is built up in the QGP phase. The figure also shows that the lifetime of the QGP phase is approximately
40\% longer at the LHC. The inclusion of viscosity does not have a large effect on the temperature evolution at the center of the LHC fireball.

These findings are in line with earlier calculations using 2+1D ideal hydrodynamics \cite{Kestin:2008bh}, 
where a small decrease of $v_2$ as a function of $p_T$ was predicted, when going to higher initial energy densities.

\begin{figure}[tb]
   \begin{center}
     \includegraphics[width=8.5cm]{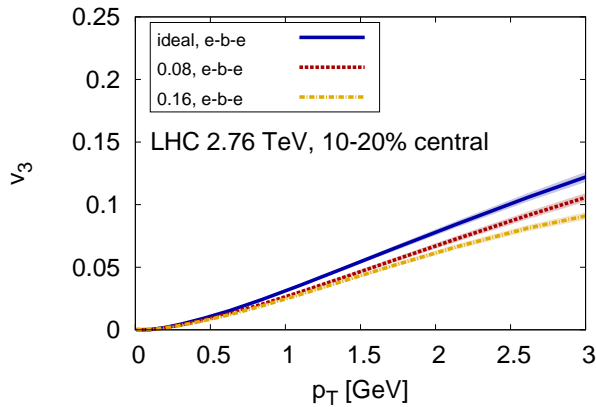}
     \caption{(Color online) Charged hadron $v_3$ in 10-20\% central collisions.}
     \label{fig:v3-10-20-new}
   \end{center}
\end{figure}

\begin{figure}[tb]
   \begin{center}
     \includegraphics[width=8.5cm]{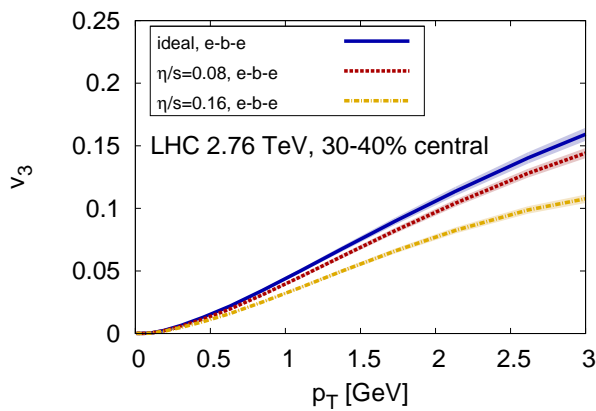}
     \caption{(Color online) Charged hadron $v_3$ in 30-40\% central collisions.}
     \label{fig:v3-30-40-new}
   \end{center}
\end{figure}

Finally, we present predictions for the triangular flow coefficient $v_3(p_T)$ for LHC energies in Figs.\,\ref{fig:v3-10-20-new} and \ref{fig:v3-30-40-new}.
The results for $\eta/s=0.08$ are again remarkably similar to those at RHIC energies \cite{Schenke:2010rr}, while the difference between curves computed with different $\eta/s$ is smaller than at RHIC energies.

We have presented first event-by-event calculations of elliptic and triangular flow at LHC energies of $\sqrt{s}=2.76\,{\rm TeV}$.
The data seems to be best described with a viscosity to entropy density ratio of $\eta/s=0.08$ or smaller. A value of $\eta/s=0.08$ was also found to describe RHIC data best earlier \cite{Schenke:2010rr}.
Generally, the Monte-Carlo Glauber initial conditions tend to lead to relatively small $v_2$ in the 3+1D event-by-event simulation compared to experimental data from ALICE \cite{Aamodt:2010pa}.
It was found that, in spite of the longer lifetime of the system at LHC, the anisotropic flow is not much larger than at RHIC: it is built up almost entirely within the first $5-6\,{\rm fm}/c$, and at the LHC the system even acquires a negative eccentricity for the 
last $5\,{\rm fm}/c$ of the evolution, reducing the momentum anisotropy. The low value of $\eta/s$ extracted from RHIC has been interpreted as the formation of a ``perfect liquid''; it would seem this description is also appropriate at the LHC. A detailed combined comparison of experimental spectra and higher harmonics using the techniques presented in this work will be able to determine the initial conditions and their fluctuations as well as the shear viscosity to entropy density ratio.


\subsection{Acknowledgments}
B.P.S. thanks Adrian Dumitru and Raju Venugopalan for fruitful discussions. We thank R. Snellings for providing the experimental data. 
This work was supported in part by the Natural Sciences and Engineering Research Council of Canada.
B.P.S. was supported in part by the US Department of Energy under DOE Contract No.DE-AC02-98CH10886, and 
by a Lab Directed Research and De\-velopment Grant from Brookhaven Science Associates.
\bibliography{hydro}

\end{document}